\documentclass[12pt]{article}
\usepackage{amssymb}
\begin{document}

\begin{titlepage}

                            \begin{center}
                            \vspace*{2cm}
        \large\bf A stochastic derivation of the geodesic rule\\

                            \vfill

              \normalsize\sf NIKOS \ KALOGEROPOULOS\\

                            \vspace{0.2cm}

 \normalsize\sf Department of Science\\
 BMCC - The City University of New York,\\
 199 Chambers St., \ New York, NY 10007, \ USA\\

                            \end{center}

                            \vfill

                     \centerline{\normalsize\bf Abstract}
                            \vspace{0.3cm}
\normalsize\rm\noindent \setlength{\baselineskip}{18pt} We argue
that the geodesic rule, for global defects, is a consequence of
the randomness of the values of the Goldstone field $\phi$ in each
causally connected volume. As these volumes collide and
coalescence, $\phi$ evolves by performing a random walk on the
vacuum manifold $\mathcal{M}$. We derive a Fokker-Planck equation
that describes the continuum limit of this process. Its
fundamental solution is the heat kernel on $\mathcal{M}$,
whose leading asymptotic behavior establishes the geodesic rule.\\

                             \vfill

\noindent\sf PACS: 02.50.Ey, 11.10.Ef, 11.30.Fs\\
    Keywords: Geodesic rule, Global defects.\\

                             \vfill

\noindent\rule{7.5cm}{0.2mm}\\
\begin{tabular}{ll}
\noindent\small\rm E-mail: & \small\rm nkalogeropoulos@bmcc.cuny.edu\\
                  & \small\rm nkaloger@yahoo.com
\end{tabular}
\end{titlepage}

%%%%%%%%%%%%%%%%%%%%%%%%%%%%%%%%%%%%%%%%%%%%%%%%%%%%%%%%%%%%%%%%%%%%%%%%%%%%%

                             \newpage

\normalsize\rm \setlength{\baselineskip}{18pt}

              \centerline{I. \ INTRODUCTION}

                            \vspace{0.3cm}

To predict the density of topological defects formed after a phase
transition, Kibble [1] postulated two ``rules": the randomness and
independence of the value of the Goldstone scalars $\phi$,
parametrize the vacuum manifold $\mathcal{M}$ within each causally
connected region, and the ``geodesic rule". According to the
latter, when two regions of $\mathcal{M}$ having different values
of  $\phi$ collide with each other, $\phi$ interpolates between
these values across the common boundary, following the shortest
path [1]. The geodesic rule is simple, and as such, aesthetically
appealing and favorable for numerical implementation, but it lacks
a thorough justification (see however [2],[3]). Moreover, there is
no consensus on its validity, or
even on its extensions in the case of local defect formation [4]-[8]. \\

We argue that the geodesic rule is a result of the stochasticity
of the selection of $\phi$ within each causal horizon. Our
argument relies on the random nature  of the collisions between
causally unrelated regions on $\mathcal{M}$. It differs from the
other treatments on this issue, which rely on energetic
considerations [2],[3]. The present discussion applies, as is, to
theories with broken global symmetries only. The presence of gauge
symmetries would introduce several subtleties in the present
arguments. These difficulties are directly related to the
fundamental lack of understanding of the vacuum structure of gauge
theories. Explicit understanding of the topology and geometry as
well as parametrizations of the underlying moduli space of the
affine connections of gauge theories is still lacking in most
cases of physical interest. As a result, even the formulation of
appropriate stochastic equations, analogous to the Fokker-Planck,
becomes problematic. For these
reasons we will deal only with the case of global defects in this treatment.\\

The organization of this paper is as follows: Section II presents
the derivation of the geodesic rule and the assumptions on which
this derivation relies. Section III analyzes the domain of
applicability of this result and  re-examines its assumptions.
Section IV presents a brief discussion and our conclusions.\\

                              \vspace{0.5cm}

%%%%%%%%%%%%%%%%%%%%%%%%%%%%%%%%%%%%%%%%%%%%%%%%%%%%%%%%%%%%%%%%%%%%%%%%%

         \centerline{II. \ DERIVATION OF THE GEODESIC RULE}

                              \vspace{0.3cm}

At the outset of our treatment, it is worth pointing out that the
Goldstone field (order parameter) $\phi$ is scalar only as far as
its space-time properties are concerned. In a general classical
field theory, $\phi$ is a section of a vector bundle $\mathcal{P}$
over a space-time $M$ which is most frequently considered to be
$\mathbb{R}^n$. The typical fiber of $\mathcal{P}$ is a linear
representation, in most cases, of a Lie group $G$ [9]. Then $\phi$
provides a local parametrization of $G$ in $\mathcal{P}$. In a
quantum or thermal field theory we start with an appropriate space
of maps $f:M\rightarrow G$, or, in this language, with an
appropriate space of sections of $\mathcal{P}$. These sections
must have some physically acceptable regularity properties, most
frequently pertaining to their differentiability, as well as
appropriate normalization properties with respect to the symmetric
or hermitian inner product of $G$ and the Riemannian inner product
of $M$ [9]. Moreover, in order to ensure the finiteness of the
energy of the system, we impose some asymptotic requirements on
$f$ which amount, very frequently, to effectively considering a
compactification of $M$. Such maps tend to form, very  frequently,
 Sobolev spaces on the sections of $\mathcal{P}$.\\

A classical/non-thermal field theory can be determined by its
Lagrangian $\mathcal{L}$ [9],[10]. A quantum/thermal field theory
can be defined by a path integral on appropriately defined
functional spaces, as was mentioned above. The result of the
calculation of such a path integral can be encoded, under some
rather general conditions, in such a way as to give rise to extra
terms to the classical potential $V_{c}$. The requirement of
renormalizability [9],[10], when applicable, often imposes
additional constraints on the nature of these extra terms, thus
severely constraining the form of the initially allowed form of
$\mathcal{L}$. Our arguments rely on the existence of such an
effective potential $V_{eff}$ [10] for the description of the
thermal field theory. Renormalizability constraints of the
underlying theory are of no concern here, since we are dealing
with effective, as opposed to fundamental, theories from the
outset. Then, the effective potential allows us to describe the
system in an apparently non-thermal manner and that
is why many of our arguments have a very strong classical/non-thermal flavor. \\

For concreteness, we begin by considering a first order phase
transition proceeding by bubble nucleation [11]. This line of
argument essentially holds also for continuous transitions
proceeding by spinodal decomposition, as we comment at the end of
the paper. We assume that the phase transition giving rise to the
topological defects proceeds by thermalization and will be
eventually completed. It is known [12],[13] that after their
formation, bubbles whose radius is larger than a critical value
$r_{crit}$ tend to expand and bubbles whose radius is smaller than
$r_{crit}$ tend to shrink until they disappear. For this reason,
we restrict our attention to bubbles of radii greater than
$r_{crit}$. We are interested in length scales much bigger than
$r_{crit}$. Then without loss of generality we can pretend that
$r$ can take any
positive value with lower bound zero.\\

A first assumption is that when two bubbles collide and coalesce,
the bubble which is generated has an effective value of $\phi$
which is the weighted average of the values of the two bubbles
that collided. In this weighted average, the contribution of each
of the two colliding bubbles should be proportional  to the number
of points it contains, i.e. to its volume. It is simplicity that
makes such a choice plausible. It is very possible that there are
systems for which such a rule for calculating the weighted average
does not apply, although we fail to see a generic reason why it
should not. Thus, if the two coalescing bubbles have values of
Goldstone fields $\phi_1$ and $\phi_2$ respectively, the resulting
bubble will have an effective value of $\phi$ given by
\begin{equation}
 \phi = \frac{V_1\phi_1 +V_2\phi_2}{V_1 + V_2}
\end{equation}
 where $V_1$ and $V_2$ are the corresponding volumes of each
bubble.\\

A second assumption is that at least two distinct, widely
separated time scales exist for such a system: $\tau_C$ which
describes the average time between two consecutive bubble
collisions and $\tau_M$ which is the mixing time of the two values
of the Goldstone fields, after the bubble collision has taken
place. These time scales are determined by the dynamics of each
model. Their specific values, important as they are in the
dynamics of phase transitions, are not relevant in our argument,
so we will not delve into the issue of their calculation any
further. This assumption is very strong as we expect that most
systems violate it, more or less. However, as we explain in
section III, for systems that follow it or weakly violate it, the
geodesic rule should still hold, due to the wide domain of
applicability of the Central Limit theorem. If $\tau_C\ll\tau_M$,
the values of $\phi$ of the two bubbles will not have enough time
to mix well between two collisions. This is a case favorable for
the formation of topological defects [14],[15]. The opposite case,
namely $\tau_C\gg\tau_M$, is very unfavorable for the formation of
defects. For this reason we examine systems that belong to the
former case. Incidentally, the probability of simultaneous
collisions of three or more bubbles is so small that the
contribution of such events to the number of produced
defects is negligible.\\

Due to the reasons given in the first two paragraphs of this
section, we need to describe aspects of the dynamics of the phase
transition as it evolves through bubble collisions on
$\mathcal{M}\times\mathbb{R}$. This evolution is parametrized by a
scalar variable $t\in\mathbb{R}$. The relation between $t$ and the
actual time in which the phase transition takes place depends on
the parameters of the specific model. We do not need to know any
further details about such a relation for our argument's sake; it
just suffices to know that such a relation exists. Because of the
random distribution of bubbles and the random way in which they
collide, $\phi$ undergoes a random evolution i.e. it undergoes a
random walk on $\mathcal{M}$. Let $P(\phi, t)$ denote the
probability density function describing this stochastic process.
To determine $P(\phi, t)$, we start by discretizing $\mathbb{R}$,
i.e. reducing it to a Euclidean one dimensional lattice,
isomorphic to $\mathbb{Z}$. We set the  spacing of this lattice to
be equal to $\tau_C$ since this is the shortest scale of $t$
associated with the evolution of bubbles. Since $\tau_M\gg\tau_C$,
we can assume without loss of generality that $\tau_M$ is an
integral multiple of $\tau_C$. Then $P(\phi, t)$ has domain
$\mathcal{M}\times\mathbb{Z}$. To explicitly indicate this
discretization, we rewrite $t$ is as
$t_n$, where $t_n\in\mathbb{Z}$.\\

A third assumption in our argument is that the evolution of the
system of bubbles described by $P(\phi,t_n)$ is a Markovian
process. This is actually an oversimplification. We have already
assumed that $\tau_M\gg\tau_C$, so after each collision the values
of $\phi$ do not mix very strongly in the resulting bubble. Then
the relative orientation of the bubbles before the collision took
place leaves its trace on the resulting bubble through the
distribution  of the values of $\phi$ on it. However, even within
our approximation, the relative orientation of two bubbles before
each collision is random. Eventually, we  average over all
possible collisions between pairs of bubbles. The averaging
process erases all trace of the relative orientation of bubbles
before a collision. In short, although the collision between two
specific bubbles would give rise to a non-Markovian process, the
randomness of orientation and the averaging process reduce the
stochastic process to a Markovian one. This stochastic process is
determined by $P(\phi, t_n)$ which obeys the Chapman-Kolmogorov
equation [16],[20]
\begin{equation}
 P(\phi_3,t_3)=\int_{\mathcal{M}}
 P(\phi_3,t_3|\phi_2,t_2) \ P(\phi_2,t_2|\phi_1,t_1)  \ d\phi_2
\end{equation}
where $P(\phi_2,t_2|\phi_1,t_1)$ denotes the conditional
probability density for the stochastic variable to have value
$\phi_2$ at $t_2$ given that it had value $\phi_1$ at $t_1$. The
definition of the probability density function $P(\phi,t_n)$ also
gives
\begin{equation}
 P(\phi,t_n+\tau_C)=\int_{\mathcal{M}} P(\phi,t_n+\tau_C|\psi,t_n) \ P(\psi,t_n) \ d\psi
\end{equation}
We are interested in the behavior of the system at large $t$
scales when compared to $\tau_C$, as the system approaches its
equilibrium, so we have to consider the continuum time limit \
$\tau_C\rightarrow 0$. Then \ $t_n$ \ is replaced by \ $t$ \ and a
Taylor series expansion with respect to $\tau_C$ gives the master
equation [16]
\begin{equation}
 \frac{\partial P(\phi ,t)}{\partial t} = \int_{\mathcal{M}} \
      \{W(\phi |\psi )P(\psi ,t)-W(\psi |\phi )P(\phi ,t)\} \ d\psi
\end{equation}
Here  \ $W(\psi |\phi )$ \ denotes the transition probability per
unit $t$ for the Goldstone field to attain the final value $\psi$,
given its initial value was \ $\phi$.\\

A considerable simplification occurs, if we further assume that
each bubble collides only with much smaller ones. This is a very
strong assumption, which we expect to be violated in may cases.
Clearly the way the bubbles evolve and collide depends on the
details of the nucleation model describing the bubbles. It is
moreover true that generally many bubbles of similar sizes collide
with each other, a fact which renders this approximation invalid.
The assumption is necessary though, in order to make the
quantitative analysis tractable. We expect that because of the
very wide range of applicability of the Central Limit theorem, the
final conclusion that we reach  applies to many cases that
initially may even violate this assumption. We  perform a Taylor
series expansion in terms of the short displacement $s$ that the
scalar field \ $\phi$ \ experiences after each collision.
Complications arise because \ $\mathcal{M}$ \ is a Riemannian
manifold which is not flat. The equations come as close as
possible to the ones of the flat case by choosing a normal
coordinate system. In this coordinate system the connection
coefficients (Christoffel symbols) are zero at the origin $O$ and,
to a first order approximation the metric  there is Euclidean. Let
$\{e_1, e_2, \ldots, e_{\dim\mathcal{M}}\}$ be a coordinate basis
of $T_o\mathcal{M}$. The master equation (4), when expanded up to
quadratic terms in $s$ in this basis [20],[21] gives
\begin{equation}
   \frac{\partial P(\phi ,t)}{\partial t} =  -\nabla_i \left[
a^i(\phi)P(\phi , t)\right] +  \frac{1}{2}\nabla_i\nabla_j \left[
         a^{ij}(\phi )P(\phi, t)\right]
 \end{equation}
where
\begin{equation}
   a^{ij}(\phi ) = \int\limits_{0}^{+\infty}  s^i s^j \ W(\phi ;s) \ ds
  \end{equation}
with
 \begin{equation}
   W(\psi|\phi) = W(\phi ;s)
    \end{equation}
where $s$ is the distance between $\phi$ and $\psi$ with respect
to the metric of $\mathcal{M}$, and $\nabla$ is a Riemannian
connection on $\mathcal{M}$, not necessarily compatible with the
metric [19]. This Fokker-Planck equation (5) has the first moment
$a^i(\phi)$ equal to zero since there are no ``external fields",
i.e. preferential directions on $\mathcal{M}$. Even if that is not
the case, and the linear term is non-zero, the stochasticity of
the directions of the bubble collisions would eliminate such a
linear term upon averaging over $T_o\mathcal{M}$. Then (5) reduces
to the diffusion equation
\begin{equation}
  \frac{\partial P(\phi ,t)}{\partial t} =
  \nabla_i\nabla_j \left[ D^{ij}(\phi )P(\phi, t) \right]
  \end{equation}
with \ $ D^{ij}(\phi ) = \frac{1}{2} \ a^{ij}(\phi )$\\

If a bubble under consideration collides with either similar in
size or much bigger bubbles, then the approximation leading to (5)
is insufficient to describe the phenomenon. In such a case we need
to keep more terms in the Kramers-Moyal expansion
 \begin{equation}
\frac{\partial P(\phi ,t)}{\partial t} =
 \sum_{n_{i_1}n_{i_2}\cdots n_{i_k}\cdots}^{\infty}
    \frac{(-1)^n}{n_{i_1}!n_{i_2}!\cdots n_{i_k}!\cdots}
      \nabla_{i_1}\nabla_{i_2}\cdots\nabla_{i_k} \cdots
            \left[ a^{i_1i_2\cdots i_k\cdots }(\phi )P (\phi, t)\right]
            \end{equation}
of the master equation (4). Actually, according to Pawula's
theorem [20], which holds for a stochastic process in Euclidean
space, the positivity of the transition probability rate, \
$W(\psi|\phi)$, \ implies that the Kramers-Moyal expansion either
terminates at the second order term (resulting in the
Fokker-Planck equation (5)) or never terminates. Considering this
theorem, our approximation leading to the termination of the
Kramers-Moyal expansion (9) at the second order term is also
optimal, in a sense, for technical reasons: any other choice would
result in infinite sub-harmonic terms which would not be
manageable without further, model-dependent, approximations. It is
worth mentioning at this point that the expansion (9) is used
heuristically. The covariant derivatives on $\mathcal{M}$ do not
commute [19], since
\begin{equation}
  \nabla_i\nabla_j \ e_k - \nabla_j\nabla_i \ e_k=R(e_i, e_j) \ e_k
\end{equation}
where  $R(e_i,e_j)e_k$ are the Riemann tensor components in the
basis of $T_o\mathcal{M}$ that we are using. The ambiguity in the
order of the covariant derivatives appearing in the terms of order
higher than two in (9), result in extra terms containing the
Riemann tensor and its covariant derivatives. The second order
term may, at worst, acquire an extra term involving the scalar
curvature of $\mathcal{M}$, a fact that does not alter the leading
dependence of the resulting solution (13) on the distance function
$d(\phi)$, as will be seen in the sequel. The order, as well as
the number of times, that the ``diffusion'' tensor components
$a^{i_1i_2\cdots i_k\cdots }(\phi )$ have to be differentiated
depend on the details of the system. These ambiguities should be
considered as an addition to the, essentially arbitrary at the
mesoscopic level, choice of the It\^{o} or the Stratonovich way of
performing the stochastic integration [16],[21], which arises in
the integration of the Fokker-Planck equation (5). In either case,
a more detailed knowledge of the microscopic behavior of the
system is required [16],[20] for the correct choices to be made.
Such a particular choice, however, plays no role in our subsequent
arguments.\\

In many cases of physical interest, the system can be described by
averaged diffusion coefficients $D^{i_1i_2\cdots i_n\cdots}$. Once
more, the averaging process over $T\mathcal{M}$ for bubble
collisions justifies this fact. Then, we can replace the different
values of the tensor components by a family of scalar functions
$\{D_l(\phi), l=1,2,\ldots\}$ in each term of (9). A class of
examples of vacua very frequently encountered in field theoretical
applications for which such a reduction is possible is provided by
homogeneous spaces [19] of compact Lie groups. Moreover, we  make
the additional assumption that the function $D(\phi)$, to which
the coefficient of (8) reduces, varies slowly over $\mathcal{M}$.
This allows us to ignore the contribution of the derivatives of
$D$ and (8) reduces to the diffusion equation
\begin{equation}
\frac{\partial P(\phi ,t)}{\partial t} =
  D\nabla^2  P(\phi, t)
\end{equation}
where $\nabla^2$ denotes the Laplace-Beltrami operator associated
with the metric of $\mathcal{M}$. Examples of a slow variation of
$D(\phi)$ are  $\mathcal{M}$ which are symmetric spaces [19]. Even
more specifically, when $\mathcal{M}$ is a compact quotient of a
sphere or a torus by the free action of a discrete group, as in
the case of projective spaces for instance, then $D(\phi)$ is
exactly constant [22].\\

 The solution of (11) with the initial condition
\begin{equation}
 P(\phi,0)=\delta(\phi)
\end{equation}
is the heat kernel of the Laplace-Beltrami operator $\nabla^2$ on
$\mathcal{M}$. Using the maximum principle one can prove that
$P(\phi,t )$, which is really the heat kernel of  \ $\nabla^2$ \
on  \ $\mathcal{M}$, is positive definite [21],[23]. Then its
normalized form can be legitimately called a probability density
function, as we have been assuming all along.  We are interested
in the behavior of the system for relatively short times $t$.
Indeed, at very long $t$ the effects of mixing will have become
significant and, as was remarked after (1), this would introduce
additional complications that could mask or even invalidate this
line of argument. At small $t$, i.e. as $t\rightarrow 0^+$ the
heat kernel has the asymptotic expansion [21],[23]
\begin{equation}
 P(\phi) = (4\pi t)^{-\frac{\dim\mathcal{M}}{2}}\exp\left(
 -\frac{[d(\phi)]^2}{4t}\right)
 \eta(d(\phi))\left\{\sum_{i=0}^k u_i(0)t^i+o(t^k)\right\}
\end{equation}
where $\eta(d(\phi))$ is a bump function which has the value one,
if $\phi$ is in the cut-locus of the origin $O$, and zero if
$\phi$ is outside it. We can then clearly see [21] that the
leading behavior of $P(\phi)$ as \ $t\rightarrow 0^+$, is
\begin{equation}
\lim_{t\rightarrow 0^+} \ t\ln P(\phi)=-\frac{[d(\phi)]^2}{2}
\end{equation}
which is actually true for any \ $\phi\in\mathcal{M}$ \
(Varadhan's asymptotic formula) [21].\\

Let the initial and final values of $\phi$ on $\mathcal{M}$ be
fixed. Then, according to (14), the stochastic process described
by (2) and (3) gives rise to a probability density $P(\phi, t)$
which is maximized if the distance function $d(\phi)$ between $O$
and $\phi$ is minimized, i.e. along the minimal geodesic
joining them. This observation is the geodesic rule.\\

                      \vspace{0.5cm}

%%%%%%%%%%%%%%%%%%%%%%%%%%%%%%%%%%%%%%%%%%%%%%%%%%%%%%%%%%%%%%%%%%%%%%%%%%%%%%%%%

\centerline{III. \ ON THE APPLICABILITY OF THE GEODESIC RULE}

                      \vspace{0.3cm}

Let $l_{min}$ and $l$ be the lengths of the minimal geodesic and
some other piecewise smooth curve of $\mathcal{M}$ joining $O$ and
$\phi$, respectively, and let $P_{min}(\phi)$ and $P(\phi)$ the
corresponding probability densities at $\phi$ in the limit
$t\rightarrow 0^+$. Then (13) implies that
\begin{equation}
 P(\phi) = P_{min}(\phi)\exp\left\{ -\frac{l^2-l_{min}^2}{4t}
 \right\}
\end{equation}
so the contribution of non-minimal paths is exponentially
suppressed in expressions like path integrals. It is worth
noticing that the geodesic rule can actually be violated. We have
argued that it is not really a rule, but rather a minimalist
prescription for calculating the density of global topological
defects. If someone is interested in calculating the number
(density) of topological defects  with some degree of accuracy
beyond a simple estimate, either analytically or numerically, the
geodesic rule will provide a fast and accurate way most of the
times. However, we expect deviations from its predictions to
become important in cases in which there are many paths for which
$P(\phi) \approx P_{min}$. One such  case occurs, if, for instance
$\mathcal{M}$ is a $C_l$ manifold [19], i.e. a compact Riemannian
manifold all the closed geodesics of which have length $l$.
Examples of $C_l$ manifolds are symmetric spaces of positive
sectional curvature, manifolds which occur frequently as vacua
of physical systems.\\

 What we have effectively done in this approach, is that we have
arrived at an integration measure on the appropriate space of
functions on $\mathcal{M}$, namely the Wiener measure on this
space [24]. This is then used to perform the thermal path integral
describing the phase transition which gives rise to the
topological defects. The path integral is essentially determined
by the Feynman-Kac formula [24], [25]
\begin{equation}
   ker \ e^{-t\mathcal{H}}=\int e^{-tV(t)dt}d\phi
\end{equation}
for the kernel of $e^{-t\mathcal{H}}$, where $\mathcal{H}$ denotes
the effective Hamiltonian on $\mathcal{M}$ of the system of
colliding bubbles. Then, in the normal coordinate system that we
are using and disregarding any curvature terms, which provide
second order corrections in $t$ and are eventually factorized in a
path integral, $V(t)$ is trivial on the path space of
$\mathcal{M}$. Therefore, the dominant contribution in the thermal
path integral comes from evaluation of the Laplacian on the
shortest possible
paths on $\mathcal{M}$, a fact which establishes the geodesic rule.\\

The result of this analysis relies on establishing, for this
system, the probability distribution given by (13). This
probability distribution is Gaussian with respect to the distance
function $d(\phi)$. The Central Limit theorem [26] states that
under some very general conditions a probability density on
$\mathcal{M}$ would converge to (13). We are then led to believe
that although (13) is derived under some rather strong
assumptions, it should be valid even in cases that violate one or
more of these assumptions, as long as the assumptions of the
Central Limit theorem are obeyed. The geodesic rule describes the
leading, universal, behavior of such systems. One class of systems
that violate the geodesic rule are ones whose bubble collisions
are described by a non-Markovian process such as, for instance, a
Levy process [17]. Even in such cases, variations of the
Generalized Central Limit theorem still hold, a fact however which
does not guarantee the validity of the geodesic rule in its
original formulation, which is the one that we have proved in this paper.\\

A case in which it is known that the geodesic rule does not hold
is when the dynamics of the order parameter is dominated by
fluctuations [4],[8],[27]. Such an observation however does not
contradict our conclusions. Indeed, when fluctuations dominate,
the absolute minimum of the effective potential [10] is what
determines $\mathcal{M}$. The loop expansion of the effective
potential, in powers of $\hbar$ or a corresponding thermal
parameter [10], shows that fluctuations force the effective
potential to develop singularities [10]. The nature of these
singularities is not clear, but it becomes amply evident through
such a procedure that $\mathcal{M}$ can no longer be considered to
be a Lie group, a homogeneous space or even a manifold as in the
classical/non-thermal case [22]. As a result, all the arguments
leading to (11) and subsequently to (13) and (15), which rely on
$\mathcal{M}$ being a Riemannian manifold do not hold any more. We
conclude then that the treatment of cases of dominant fluctuations
lies outside the scope of our approach, hence there is no
contradiction with our results.\\

                        \vspace{0.5cm}

%%%%%%%%%%%%%%%%%%%%%%%%%%%%%%%%%%%%%%%%%%%%%%%%%%%%%%%%%%%%%%%%%%%%%%%%

        \centerline{IV. \ DISCUSSION AND CONCLUSIONS}

                        \vspace{0.3cm}

In the case of a first-order phase transition [11] there is an
energy barrier that $\phi$ has to overcome in order to move from
the meta-stable to the stable vacuum. If we consider the Fourier
transform of the Fokker-Planck equation, then we expect that
$\tilde{D}(k)$ will have a strong peak on wave-numbers $k$
comparable to the width of the potential well around the
meta-stable vacuum. In the case of continuous transitions
(spinodal decomposition), there is no longer an energy barrier
that would act as a localizing factor and would force
$\tilde{D}(k)$ to have sharp peaks. The arguments presented above
do not use this localization in momentum space, therefore they are
equally valid in both cases. This, of course, does not mean that
the two cases are physically identical. In a continuous
transition, we can still consider a  region that shares the same
value of the scalar field, but does not possess a physical
boundary, like a bubble wall, as in the case of first
order phase transitions.\\

To conclude, we have shown that the geodesic rule holds for a
general class of systems undergoing a phase transition, as a
leading approximation. Violations of the rule can and will
generally occur. The validity of this rule depends on establishing
a Gaussian process through which the phase transition from the
meta-stable to the stable vacuum proceeds. For any system that can
be described by such a stochastic process, which obeys the
assumptions of the Central Limit theorem, the density of the
resulting global defects can be accurately estimated by using
the geodesic rule.\\

                                \vspace{0.1cm}

\noindent {\sc Acknowledgements:} \ We are grateful to Dr. P.
Benetatos for his constructive criticism of the manuscript. We are
also grateful to the referee for pointing out reference [27] and
for several comments that helped improve the presentation of this
paper.

%%%%%%%%%%%%%%%%%%%%%%%%%%%%%%%%%%%%%%%%%%%%%%%%%%%%%%%%%%%%%%%%%%%%%%%%%%%%%%%%%

                                \vspace{1cm}

\noindent\normalsize\centerline{\sc References}

                                \vspace{0.3cm}

\setlength{\baselineskip}{18pt} \noindent\rm
 1. T.W.B. Kibble, \emph{J. Phys.} \bf{A9}, \rm 1387 (1976)\\
 2. A.M. Srivastava, \emph{Phys. Rev.} \bf{D46}, \rm  1353 (1992)\\
 3. A.M. Srivastava, \emph{Phys. Rev.} \bf{D45}, \rm  R3304 (1992)\\
 4. E. Copeland and P. Saffin, \emph{Phys. Rev.} \bf{D54}, \rm 6088 (1996)\\
 5. M. Hindmarsh, A.C. Davis and R. Brandenberger, \emph{Phys. Rev.} \bf{D49},
                                                        \rm 1944 (1994)\\
 6. T.W.B. Kibble and A. Vilenkin, \emph{Phys. Rev.} \bf{D52}, \rm 679 (1995)\\
 7. L. Pogosian, T. Vachaspati, \emph{Phys. Lett.} \bf{B 423}, \rm 45 (1998)\\
 8. M. Hindmarsh, A. Rajantie, \emph{Phys. Rev. Lett.} \bf{85}, \rm 4660 (2000) \\
 9. P. Deligne et. al., \emph{Quantum Fields and Strings: A course
                                               for Mathematicians}, \ AMS (1999)\\
 10. S. Weinberg, \emph{The Quantum Theory of Fields}, Vols. I,II,
                                             \ Camb. Univ. Press (1995)\\
 11. A.J. Bray, \emph{Adv. Phys} \bf{43}, \rm 357 (1994)\\
     J.S. Langer, in \emph{Solids Far From Equilibrium},
                                 C. Godreche Ed., \ Camb. Univ. Press (1992)\\
 12. I.M. Lifshitz and V.V. Slyozov, \emph{J. Phys. Chem. Solids}
                       \bf{19}, \rm 35 (1961)\\
 13. C. Wagner, \emph{Z. Elektrochem.} \bf{65}, \rm 581 (1961)\\
 14. A. Vilenkin and E.P.S. Shellard, \emph{Strings and other
                              Topological Defects}, \ Camb. Univ. Press (1994)\\
 15. M. Kleman, in \emph{Formation and Interaction of Topological
                Defects}, A.C. Davis and R. Brandenberger, Eds., \ Plenum Press (1995)\\
 16. N.G. Van Kampen, \emph{Stochastic Processes in Physics and
                                         Chemistry} 2nd Ed., \ North Holland (1992)\\
 17. M.E. Schlesinger, G.M. Zaslavsky, U. Frisch, Eds.
               \emph{Levy flights and related topics in Physics}, \ Springer-Verlag (1995)\\
 18. J.P. Bouchaud, A. Georges, \emph{Phys. Rep.} {\bf 195}, 127 (1989)\\
 19. A. Besse, \emph{Einstein manifolds},  \ Springer-Verlag (1987)\\
 20. H. Risken, \emph{The Fokker-Planck equation}, 2nd Ed., \ Springer-Verlag (1996)\\
 21. E.P. Hsu, \emph{Stochastic Analysis on Manifolds}, \ AMS (2002)\\
 22. M. Gromov, \emph{Metric structures for Riemannian and
                            non-Riemannian spaces}, \ Birkhauser (2001)\\
 23. T. Aubin, \emph{Some Nonlinear Problems in Riemannian
                            Geometry}, \ Springer-Verlag (1998)\\
 24. D.W. Stroock, \emph{An Introduction to the Analysis of Paths
                            on a Riemannian Manifold}, \ AMS (2000)\\
 25. J. Glimm, A. Jaffe, \emph{Quantum Physics: A functional integral point of view},
                            2nd Ed., \ Springer-Verlag (1987)\\
     H. Kleinert, \emph{Path Integrals in Quantum Mechanics, Statistics, Polymer Physics
                            and Financial Markets}, 3rd Ed., \ World Scientific (2003)\\
 26. D.W. Stroock, \emph{Probability Theory: An analytic view}, \ Camb. Univ. Press (1999)\\
 27. S. Digal, S. Sengupta, A.M. Srivastava, \emph{Phys. Rev.}
     {\bf D54}, \rm 103510 (1998)\\

                                 \vfill

\end{document}